\documentclass[12pt,authoryear]{elsarticle}
\usepackage{amsmath, amsthm, amssymb}
\usepackage{natbib}
\usepackage[margin=2.8cm]{geometry}
\usepackage{hyperref,xcolor}
\journal{Statistics \& Probability Letters}
\newcommand{\tr}{^{\prime}}
\newcommand{\diag}{{\rm diag}}
\def\bd#1{\mbox{\boldmath $#1$}}    

\def\cg#1{\mbox{${\cal #1}$}}
\def\cgl#1{\mbox{\scriptsize {${\cal #1}$}}}
\newcommand{\bksl}{\tiny\mbox{{$\backslash$}}}
\newcommand{\bks}{\mbox{$\backslash$}}

\newtheorem{lemma}{Lemma}

\newtheorem{proposition}{Proposition}
\newtheorem{example}{Example}
\newtheorem{remark}{Remark}
\newtheorem{corollary}{Corollary}
\begin{document}

\begin{frontmatter}
\title{Marginal log-linear models and mediation analysis}

\author[A1]{Antonio Forcina \corref{mycorrespondingauthor}}
\address[A1]{Dipartimento di Economia, University of Perugia, via Pascoli 10, 06100, Italy}
\cortext[mycorrespondingauthor]{A. Forcina. forcinarosara@outlook.it: \url{https://scholar.google.it/citations?user=x4ONwSsAAAAJ&hl=it}}

\begin{abstract}
We review some not well known results about marginal log-linear models, derive some new ones and show how they might be relevant in mediation analysis within logistic regression. In particular, we elaborate on the relation between interaction parameters defined within different marginal distributions and describe an algorithm for estimating the sane interaction parameters within different marginals.
\end{abstract}


\end{frontmatter}
\section{Introduction}
\label{Sec-int}
%
Marginal log-linear models, \cite{Ruder}, were conceived to construct discrete multivariate distributions subject to restrictions imposed, simultaneously, on different marginals. Consider the simple context where $X$ denotes a treatment, $W$ one or more variables which might be affected by $X$ and may influence the response $Y$ which, for simplicity, we assume to be binary. In this context, we might be interested in the marginal distributions $XW$ and $XY$ in addition to the joint distribution $XWY$.
%
\subsection{Notations and preliminary results}
A list of variables, say $(X,W,Y)$, shortened as $XWY$ will be used to denote both a marginal distribution and the interaction among the variable in the list; let $\cg I$, $\cg M$ denote two such lists with $\cg I\subseteq\cg M$; $\bd\lambda_{\cgl I,\cgl M}$ will denote the log-linear interactions $\cg I$ defined within the marginal $\cg M$, coded either as contrasts between adjacent categories (Ac) or with respect to a reference category (Rc) depending on the context; in both cases, variables in $\cg M\backslash \cg I$ will be set to the initial or reference category coded as 0. When $X$ and $W$ quantitative, the linear logistic model including the $XW$ interaction  has the form
\begin{equation}
\label{Eq:logit}
\log\frac{P(Y=1\mid X=x,W=w)}{P(Y=0\mid X=x,W=w)} = \beta_0+\beta_X x+\beta_W w + \beta_{XW} xw
\end{equation}
where the $XY$ interaction is equal to $\beta_X$ under Ac and to $\beta_X x$ under Rc.

To introduce the mixed parametrization, recall that in a general multi-way table with $k$ cells, the saturated model may be parameterized as
\begin{equation}
\label{Eq:loglin}
\bd p =\bd G\bd\theta-\bd 1_k\log[\bd 1_k\tr\exp(\bd G\bd\theta)],
\end{equation}
where $\bd G$ is made of $k-1$ linearly independent columns which do not span the unitary vector and $\bd\theta$ is a vector of log-linear (canonical) parameters. Let $\bd H$ be the left inverse of $\bd G$ such that $\bd H\bd 1_k$ = $\bd 0$, then (\ref{Eq:loglin}) may be inverted as $\bd\theta$ = $\bd H\log \bd p$; note the one to one correspondence between the rows of $\bd H$, the columns of $\bd G$ and the log-linear parameters. Define the vector of mean parameters $\bd\mu$ = $\bd G\tr\bd p$; clearly there is a one to one correspondence between elements of $\bd\mu$ and $\bd\theta$. Let $\bd G_{\cgl I}$ be the collection of columns of $\bd G$ that correspond to the set of interactions in $\cg I$, then the vector $\bd\mu_{\cgl I}$ = $\bd G_{\cgl I}\tr\bd p$ has the same size as $\bd\theta_{\cgl I}$.

Given a partition of the collection of all possible interactions for the joint distribution into two disjoint sets $\cg U,\:\cg V$, the mixed parametrization, \cite[][pag. 121-22]{Barndorff1978}, is made of $(\bd\mu_{\cgl V},\bd\theta_{\cgl U})$ and has the following properties:
\begin{lemma} \label{Bar-Nil}
(i) there is a one to one mapping between $\bd p$ and $(\bd\mu_{\cgl V},\bd\theta_{\cgl U})$, (ii) the two components of the mixed parametrization are variation independent and (iii) the expected information matrix is block diagonal.
\end{lemma}
The following results on the differential properties of the mixed parametrization will be used later: let $\cg I\subseteq \cg M$ and $\bd p_{\cgl M}$ denote the distribution within the marginal $\cg M$; let $\bd\Omega(\bd p)$ = $\diag(\bd p)-\bd p\bd p\tr$, then we have \cite[see][Lemma 3, 4]{Forcina2012};
\begin{lemma} \label{Le:Cov}
$$
\bd C_{\cgl I} = \frac{\partial \bd\mu_{\cgl I}}{\partial \bd\theta_{\cgl I}\tr} =
\frac{\partial \bd\mu_{\cgl I}}{\partial \bd\eta_{\cgl I,\cgl M}\tr} =
G_{\cgl I}\tr \bd\Omega(\bd p)\bd G_{\cgl I};
$$
in addition, $\bd C_{\cgl I}$ is symmetric and positive definite if the elements of $\bd p$ are strictly positive.
\end{lemma}
\section{Main results}
\label{Sec:Main}
It is well known that the parameters in the marginal logistic models for $Y|W=w$, $Y\mid X=x$ and $W\mid X=x$ do not determine those in  (\ref{Eq:logit}); the mixed parametrization allows to sharpen this result as follows:
\begin{proposition}
(i) The parameters of the three logistic regression models defined on the marginals $XW$, $XY$, $WY$ are variation independent from $\beta_{XW}$. (ii) If $\beta_{XW}=0$, then the parameters of the three marginals determine uniquely the joint distribution.
\end{proposition}
{\sc Proof}: the log linear parameters within the $XW,\:WY,\:XY$ marginals are uniquely determined by the set of mean parameters $\bd\mu_X,\:\bd\mu_W,\:\bd\mu_Y,\:\bd\mu_{XW},\:\bd\mu_{XY},\: \bd\mu_{WY}$ which are variation independent from $\bd\theta_{XWY}$ = $\bd\beta_{XW}$. The above list of mean parameters together with $\bd\theta_{XWY}$ constitute a mixed parametrization of the joint distribution, thus (ii) follows from Lemma \ref{Bar-Nil}. $\Box$
\begin{remark}
In principle, under (ii), the parameters in (\ref{Eq:logit}) could be written as functions of the mean parameters; the algorithm in \cite{Forcina2012}, A2, provides an efficient and accurate numerical alternative.
\end{remark}

For the model in (\ref{Eq:logit}), \cite{Stangh2019marginal} derived an expression for $\delta_X$ = $\beta_X-\beta_X^{*}$, where $\beta_X^{*}$ is the regression coefficient of $X$ in the linear logistic model defined within the marginal $XY$ distribution.  For the case of a multivariate discrete distribution on a set of binary random variables, an expression for the difference between the same interaction parameters defined within two different marginals, say $\cg N\subset\cg M$, was derived by \cite{evans2015smoothness}, Theorem 3.1. In the Appendix we rewrite the latter result in the case where $X$ and $W$ are discrete and show that, by setting $\cg N=XY$ and $\cg M=XWY$, they are essentially equivalent to those in \cite{Stangh2019marginal}.

The following provides some additional insights into the relation between interaction parameters defined within different marginals:
\begin{proposition} \label{Pr:p1}
Suppose that $\bd\theta_{\cgl I}$ has size $d$, then
\begin{equation}
\label{Eq:Deriv}
\frac{\partial\bd\lambda_{\cgl I,\cgl M}}{\partial\bd\theta_{\cgl I}\tr} = \frac{\partial\bd\lambda_{\cgl I,\cgl M}}{\partial\bd\mu_{\cgl I}\tr}
\frac{\partial\bd\mu_{\cgl I}}{\partial\bd\theta_{\cgl I}\tr} = \bd I_d.
\end{equation}
\end{proposition}
{\sc Proof}: Follows from Lemma \ref{Le:Cov}. $\Box$

In the special case when $d=1$, Proposition \ref{Pr:p1} simply says that $\beta_X-\beta_X^{*}$ and $\beta_X$ are variation independent which is somehow implied by the derivation in \cite{Stangh2019marginal}. Additional features of the result are clarified in the example below.
\begin{example}
Consider an $XWY$ distribution where $W,\:Y$ are binary and $X$ has $k$ categories; suppose we have two probability distributions $\bd p^1,\:\bd p^2$, with all log-linear parameters being equal, except for $\bd\theta_{XY}^1\neq\bd\theta_{XY}^2$. Then the difference between corresponding pairs of marginal interactions $\bd\lambda_{XY,XY}^1- \bd\lambda_{XY,XY}^2$ is equal to $\bd\theta_{XY}^1-\bd\theta_{XY}^2$.
\end{example}

It is well known that we cannot impose log-linear restrictions on the $XY$ interactions both in the $XY$ marginal and in the joint distribution; for a formal argument see \cite{Ruder}.
However, \cite{CoFor2014} proved a result that, within the Rc coding and assuming that $W$ has $m$ categories, may be stated as follows:
\begin{proposition} \label{Pr:p2}
Within $XWY$, the marginal log-linear parametrization with elements
$$
(\bd\lambda_{X,XY},\:\bd\lambda_{Y,XY},\:\bd\lambda_{XY,XY},\:\bd\lambda_{W,XW},\:\bd\lambda_{XW,XW},\:
\bd\theta_{XY},\:\bd\theta_{WY},\:\bar{\bd\theta}_{XWY})
$$
where $\bar{\bd\theta}_{XWY}$ is obtained from $\bd\theta_{XWY}$  by deleting all elements with $W=w\ne 0$ is a smooth parametrization of the saturated model.
\end{proposition}
In words, if we want to define (and possibly constraint) the $XY$ interactions both in the marginal and in the joint, we need to remove a subset of the $XWY$ interactions corresponding to a fixed value of $W$. This may be seen as an added flexibility in the modelling process: if we are interested in imposing constraints to the $XY$ interaction both in the marginal $XY$ and in the joint, the price to pay is that we cannot model a subset of the $XWY$ interactions. The feature is illustrated in the next section.
\section{Application}
\subsection{The data}
The data come from the NCDS, a UK cohort study that included everybody born in UK from March 3rd to March 9th 1958. several variables concerning the parents and the child are recorded; a full description of the data set is available at
\url{http://cls.ucl.ac.uk/cls-studies/1958-national-child-development-study}.
In this simplified analysis, we consider the number of years of schooling for each parent, parents' concern about the education of the child shown at different stages (as recorded by the teachers), the weekly income of parents and the academic qualification reached by the child, an ordered categorical variable with four categories. The issue of interest is the effect of parents' education on that of the child.
Intuitively, parents' education might affect income by which to offer better chances to the child. In addition, more educated parents might show more concern being more aware of the importance of education. Direct effects may work through the atmosphere inside the family, like having books and meeting more educated friends.

For simplicity, the analysis below is restricted to the sample of 2161 daughters, the response $Y=1$ if the child got at least an high school degree; income and concern are dichotomized at the median. The exposure $X$ is a categorical variable with four levels obtained by splitting at quantiles the following measure of parent's education
$$
\tilde X = E_m+E_f-\mid E_m-E_f\mid /3,
$$
where $E_m,E_f$ denote the number of years of schooling for mother and father and $\mid E_m-E_f\mid /3$ is a penalty for unequally educated parents. We also assume there are two mediators: $U$, the father weekly income (that of the mother wa ignored, having a large number of missing values) and $V$, an average measure of the concern shown by parent at different stages, as recorded by teachers. Finally define $W$ = $(U,V)$.
\subsection{Two alternative models}
We compare two alternative models, both parameterized with the adjacent coding; because all variables except X are binary, assuming that, say, the $YX$ adjacent interactions are constant in $X$ is equivalent to assume that the logits of $Y\mid X=x$ is a linear functions of $x$. However, because the evidence against linearity in $X$  was rather strong, the dependence on $X$ was left unconstrained.
\begin{description}
\item{M1:} Define the overall effect of $X$ on $Y$ in the corresponding marginal distribution, in addition, model the effect of $X$ on the mediators in the marginal $XUV$. Define all other interactions within the joint $XUVY$, including the $XY$ interactions; the parameters already in the model determine the $XUVY$ interactions which cannot be modeled. Then we constrain to 0 the $XUV$ interactions in $XUV$ and the $YUV$ and $XVY$ interactions within $XUVY$; this model fits well with a deviance of 7.82 and 7 dof. Parameter estimates and standard errors for interaction parameters involving the $XY$ term are given in Table 1.
\begin{table}[h!]
\caption{
\label{table-T1}
Estimates of interactions containing the $XY$ in the M1 and M2 models.}
\vskip1mm
\centering
\begin{tabular}{lrrrrrr} \hline
\multicolumn{7}{c}{Estimates under M1} \\
 & \multicolumn{2}{c}{$\lambda_{XY,XY}$} & \multicolumn{2}{c}{$\theta_{XY}$} &
 \multicolumn{2}{c}{$\theta_{XYU}$} \\
$X$ & Est. & s.e. &  Est. & s.e. & Est. & s.e. \\ \hline
$0\rightarrow 1$ & -0.0066 &  0.1016 & -0.1948 &  0.2311 &  0.6587 &  0.1672 \\
$1\rightarrow 2$ &  0.5990 &  0.3194 &  0.7854 &  0.1230 & -1.5142 &  0.2817 \\
$2\rightarrow 3$ &  1.2045 &  0.1339 &  0.8004 &  0.1827 &  0.4604 &  0.1215 \\
\hline
\multicolumn{7}{c}{Estimates under M2}\\
 & \multicolumn{2}{c}{$\theta_{XY}$} & \multicolumn{2}{c}{$\theta_{XYV}$} &
 \multicolumn{2}{c}{$\theta_{XYU}$} \\
$X$ & Est. & s.e. &  Est. & s.e. & Est. & s.e. \\ \hline
$0\rightarrow 1$ &  -0.0441  & 0.1035 & -0.3048 &  0.1729 &  0.3282 &  0.2287 \\
$1\rightarrow 2$ &   0.6216  & 0.2649 &  0.3186 &  0.2581 & -0.6978 &  0.1236\\
$2\rightarrow 3$ &   0.7761  & 0.1346 &  0.0240 &  0.1182 &  0.0964 &  0.1807\\
\hline
\end{tabular}
\end{table}
\item{M2:} Define the effects of $X$ on $U,V$ within the $XUV$ marginal as above and all other effects within the joint $XUVY$; next,  constrain to 0 the $XUV$ interactions in the $XUV$ marginal as above and the $UVY$ and $XUVY$ interactions in the joint. This model, which is the closest analog to the one considered above, has a deviance of 13.04 with the same number of dof. Estimates and standard errors for the dependence of $Y$ on $X$ are displayed in Table 1.
\begin{table}[h!]
\caption{
\label{table-T2}
Model M1: dependence of income, $U$, and concern, $V$, on parents' education.}
\vskip1mm
\centering
\begin{tabular}{lrrrr} \hline
 & \multicolumn{2}{c}{$XU$ } & \multicolumn{2}{c}{$XV$}  \\
$X$ & Est. & s.e. &  Est. & s.e.  \\
\hline
$0\rightarrow 1$ &  0.4470  & 0.3175 & -0.0718 &  0.1591\\
$1\rightarrow 2$ &  0.2616  & 0.1298 &  0.4766 &  0.1478\\
$2\rightarrow 3$ &  0.9327  & 0.1762 &  1.2654 &  0.0928\\
 \hline
\end{tabular}
\end{table}
\end{description}
The effect of $X$ on $U,V$ is strongest in going from 2 to 3; the same holds for the marginal effect of $X$ on $Y$. Within M2 the effects of $X$ conditional on $U=V=0$ and $U=1,V=0$ are roughly similar the the corresponding ones under M1.

If we assume that there are no unobserved confounders, the estimated joint distribution under M1 allows to compute an estimate of the natural direct and indirect effect of parents' education on academic qualification of the daughter, \cite{Pearl2014Mediation}, by changing $X$ from one category to the next \cite[see][equations (1) and (2)]{VanderWeele2013}. Results are in Table \ref{table-T3} with standard errors estimated by bootstrap; the direct effect is always the largest component of the total though going from 0 to 1 does not seem to matter.
\begin{table}[h!]
\caption{
\label{table-T3}
Natural direct and indirect effects, when changing parents' education from one category to the next, on that of their daughters}
\vskip1mm
\centering
\begin{tabular}{lrrrrrr} \hline
 & \multicolumn{2}{c}{$0\rightarrow 1$} & \multicolumn{2}{c}{$1\rightarrow 2$} &
 \multicolumn{2}{c}{$2\rightarrow 3$}\\
 & Est & s.e.  & Est & s.e. & Est & s.e. \\ \hline
Dir.  &  -0.0107 &  0.0196 &  0.0609 &  0.0255 &  0.1549 &  0.0298\\
Ind.  &   0.0083 &  0.0069 &  0.0298 &  0.0100 &  0.1092 &  0.0179\\
Total &  -0.0024 &  0.0210 &  0.0907 &  0.0257 &  0.2641 &  0.0299\\
\hline
\end{tabular}
\end{table}
\section*{Appendix}
\subsection*{Rephrasing Robin Evans result}
Let $\cg N\subset \cg M$ be two nested marginals and $\cg R$ = $\cg M\bks \cg N$; assume that we define interactions as contrasts relative to the reference category coded as 0; we also use the convention that, when the value of the conditioning variables are not given, they are fixed to the reference value; the derivation below is, essentially, a re-writing of \cite{evans2015smoothness}. Let $\lambda_{\cgl I;\cgl K}(\bd x_{\cgl I})$ denote the log-linear interaction among variables in $\cg I$ computed within the marginal $\cg K$ fixed at the value $\bd x_{\cgl I}$.
\begin{lemma} \label{E1}
\begin{equation}
\lambda_{\cgl I;\cgl M}(\bd x_{\cgl I}) - \lambda_{\cgl I;\cgl N}(\bd x_{\cgl I}) = \sum_{\cgl J\subseteq \cgl I}(-1)^{|\cgl I\bksl \cg J|} \log p_{\cgl R\mid \cgl N}(\bd 0_{\cgl R}, \bd x_{\cgl J};\bd 0_{\cgl N\bksl \cgl J}),
\end{equation}
where the conditional probabilities on the right-hand side are of the event $\bd x_{\cgl R}=\bd 0_{\cgl R}$ when the conditioning set is split into a component taking the original values and the remaining ones fixed to 0.
\end{lemma}
{\sc Proof}: Start from the expansion of $\lambda_{\cgl I;\cgl M}(\bd x_{\cgl I})$, add and subtract $\lambda_{\cgl I;\cgl N}(\bd x_{\cgl I})$ and write the difference between the two in terms of conditional probabilities
\begin{align*}
\lambda_{\cgl I;\cgl M}(\bd x_I)
& = \sum_{\cgl J\subseteq \cgl I} (-1)^{|\cgl I\bksl \cgl J|}\log p_{\cgl M}(\bd x_{\cgl J},\bd 0_{\cgl M\bksl \cgl J}) \\
& = \sum_{\cgl J\subseteq \cgl I}(-1)^{|\cgl I\bksl \cgl J|}\log\frac{p_{\cgl M}(\bd x_{\cgl J},\bd 0_{\cgl N\bksl \cgl J}, \bd 0_{\cgl R})} {p_{\cgl N}(\bd x_{\cgl J},\bd 0_{\cgl N\bksl \cgl J})} + \lambda_{\cgl I;\cgl N}(\bd x_{\cgl I}).
\end{align*} $\Box$

We now apply Lemma \ref{E1} to the special case where $\cg M$ = $XWY$, $\cg N$ = $XY$, $Y$ is binary and $X,\:W$ are discrete; to simplify notations, let $p_{\bar W}(x,y)$ = $P(W=0\mid X=x,Y=y)$; in addition, because $XWY$ is the joint distribution, replace $\lambda_{\cgl I;\cgl M}$ with $\theta_{\cgl I}$.
\begin{corollary} \label{C1}
\begin{equation}
\lambda_{XY}(x,y) - \lambda_{XY;XY}(x,y) = \log\frac{p_{\bar W}(0,0)\: p_{\bar W}(x,y)}
{p_{\bar W}(0,y) \:p_{\bar W}(x,0)}
\end{equation}
this may also be expressed in terms of log-linear parameters defined within $XWY$ as
\begin{align*}
& \lambda_{XY}(x,y) - \lambda_{XY;XY}(x,y) = -\log\frac{ 1+\sum_{w>0}\exp[\lambda_W(w)]}
{1+\sum_{w>0}\exp[\lambda_W(w) +\lambda_{WX}(w,x)]}\\
&+\log\frac{1+\sum_{w>0}\exp[\lambda_W(w)+\lambda_{WY}(w,y)]}
{1+\sum_{w>0}\exp[\lambda_W(w)+\lambda_{WX}(w,x)+\lambda_{WY}(w,y)+\lambda_{WXY}(w,x,y)]}
\end{align*}
\end{corollary}
{\sc Proof}. The first part follows from Lemma \ref{E1} by noting that, because $\cg I$ has just two elements, the expansion contains four elements which can be arranged into the form of a log odds ratio. For the second part, first write the conditional distribution of $W\mid X,Y$ as a multinomial and then apply (3) in \cite{CoFor2014} for expanding interactions conditional to $X,Y$ into a sum of higher order interactions.
%
\subsection*{Log-linear versus logistic parameterizations}
For what follows, it might be useful to recall how, under the corner point coding, log-linear parameters may be mapped into the corresponding logistic parameters. When the dependent variable, like $Y$, is binary, we have
$$
\log\frac{P(Y=1\mid X=x,W=w)}{P(Y=0\mid X=x,W=w)} = \lambda_Y+\lambda_{XY}(x)+\lambda_{WY}(w)+\lambda_{XWY}(x,w),
$$
with the convention that the log.linear parameter is 0 whenever at least one of the arguments is 0.
Having assumed that $W$ is multinomial with, possibly, more than two categories, its logits may be written as
$$
\log\frac{P(W=w\mid X=x,Y=y)}{P(W=0\mid X=x,Y=y)} = \lambda_W(w)+\lambda_{XW}(w,x)
+\lambda_{WY}(w,y)+\lambda_{XWY}(x,w,y).
$$
\subsection*{The results of Stanghellini and Doretti}
As above, let $Y$ be binary and $X,\:W$ be discrete; equation (A2) in \cite{Stangh2019marginal} may be written as
$$
\log\frac{P(W=w\mid Y=1,X=x)}{P(W=w\mid Y=0,X=x)} = \log\frac{P(Y=1\mid X=x,W=w)}{P(Y=0\mid X=x,W=w)} - \log\frac{P(Y=1\mid X=x)}{P(Y=0\mid X=x)}
$$
which follows by expanding the left-hand side as
$$
\log\frac{P(W=w\mid Y=1,X=x)}{P(W=w\mid Y=0,X=x)} = \log\frac{P(W=w, Y=1,X=x)}{P(W=w, Y=0,X=x)} -
\log\frac{P(Y=1,X=x)}{P(Y=0,X=x)}
$$
and noting that logits may be computed equivalently either on the joint or conditional distribution.

To derive an extension of their (A3) to non binary $W$, first swap conditioning
$$
\log\frac{P(W=w\mid X=x,Y=y)}{P(W=0\mid X=x,Y=y)} = \log\frac{P(Y=y\mid X=x,W=w)}{P(Y=y\mid X=x,W=0)} +
\log\frac{P(W=w\mid X=x)}{P(W=0\mid X=x)}
$$
next expand the first term on the right hand-side by adding and subtracting $\log P(Y=0\mid X=x,W=w)$ and
$\log P(Y=0\mid X=x,W=0)$,
\begin{align*}
& \log\frac{P(Y=y\mid X=x,W=w)}{P(Y=y\mid X=x,W=0)} = \log\frac{P(Y=0\mid X=x,W=w)}{P(Y=0\mid X=x,W=0)}+ \\
& y\left[\log\frac{P(Y=1\mid X=x,W=w)}{P(Y=0\mid X=x,W=w)}-\log\frac{P(Y=1\mid X=x,W=0)}{P(Y=0\mid X=x,W=0)}\right].
\end{align*}
Thus the analog of the log-linear expansion in their (A3) is
\begin{align*}
& \log\frac{P(W=w\mid X=x,Y=y)}{P(W=0\mid X=x,Y=y)} = +y\left[\lambda_{WY}(w)+\lambda_{XWY}(x,w)\right]\\
&+\log\frac{1+\exp(\lambda_Y+\lambda_{XY}(x))} {1+\exp(\lambda_Y+\lambda_{XY}(x)+\lambda_{WY}(w)+
\lambda_{XWY}(x,w))}+\lambda_W(w)+\lambda_{WX}(w,x).
\end{align*}

which is a equivalent to (\ref{C1}) in the special case when $W,\:Y$ are both binary variables

\subsection{Acknowledgments}
This research did not receive any specific grant from funding agencies in the public, commercial, or
not-for-profit sectors. The author would like to thank Elena Stanghellini for suggesting the problem and for several helpful comments.
 \bibliographystyle{elsarticle-harv}
 \bibliography{MLL}
\end{document}